\documentclass[conference]{IEEEtran}
\IEEEoverridecommandlockouts

\makeatletter
\let\@fnsymbol\dagger
\makeatother

\usepackage{cite}
\usepackage{amsmath,amssymb,amsfonts}
\usepackage{algorithmic, listings}
\usepackage{graphicx}
\usepackage{textcomp, url}
\usepackage{xcolor, booktabs, xspace}
\usepackage{comment, threeparttable, tabularx, array}
\usepackage{color}  
\usepackage{inconsolata}  

\newcommand{\myparagraph}[1]{\vspace{0.25em}\noindent\textbf{#1:}}
\let\paragraph=\myparagraph 

\definecolor{codegreen}{rgb}{0,0.6,0}
\definecolor{codegray}{rgb}{0.6,0.6,0.6}
\definecolor{codepurple}{rgb}{0.58,0,0.82}
\definecolor{backcolour}{rgb}{1,1,1}  

\lstdefinestyle{editorstyle}{
    backgroundcolor=\color{backcolour},
    commentstyle=\color{codegreen},
    keywordstyle=\color{blue},
    numberstyle=\tiny\color{codegray},
    stringstyle=\color{codepurple},
    basicstyle=\ttfamily\fontsize{7pt}{8pt}\selectfont,  
    breakatwhitespace=false,         
    breaklines=true,                 
    captionpos=b,                    
    keepspaces=true,  
    numbers=left,                    
    numbersep=5pt,                  
    showspaces=false,                
    showstringspaces=false,
    showtabs=false,                  
    tabsize=2,  
    lineskip={-1.5pt},  
    aboveskip={0pt},
    belowskip={0pt}
}

\newcommand{\tool}{PathSentinel\xspace}

\def\BibTeX{{\rm B\kern-.05em{\sc i\kern-.025em b}\kern-.08em
    T\kern-.1667em\lower.7ex\hbox{E}\kern-.125emX}}

\begin{document}

\title{Static Detection of Filesystem Vulnerabilities \\ in Android Systems}

\author{

\IEEEauthorblockN{Yu-Tsung Lee}
\IEEEauthorblockA{\textit{Penn State University}\\
\texttt{yxl74@psu.edu}}

\and
\IEEEauthorblockN{Hayawardh Vijayakumar}
\IEEEauthorblockA{\textit{Samsung Resaerch America}\\
\texttt{h.vijayakuma@samsung.com}}

\and
\IEEEauthorblockN{Zhiyun Qian}
\IEEEauthorblockA{\textit{UC Riverside}\\
\texttt{zhiyunq@cs.ucr.edu}}

\and
\IEEEauthorblockN{Trent Jaeger$^{\dagger}$\thanks{$\dagger$~More authors to be added pending corporate approval.}}
\IEEEauthorblockA{\textit{UC Riverside}\\
\texttt{trentj@cs.ucr.edu}}


}

\maketitle

\begin{abstract}
Filesystem vulnerabilities persist as a significant threat to Android systems, despite various proposed defenses and testing techniques. The complexity of program behaviors and access control mechanisms in Android systems makes it challenging to effectively identify these vulnerabilities. In this paper, we present \tool, which overcomes the limitations of previous techniques by combining static program analysis and access control policy analysis to detect three types of filesystem vulnerabilities: path traversals, hijacking vulnerabilities, and luring vulnerabilities. By unifying program and access control policy analysis, \tool identifies attack surfaces accurately and prunes many impractical attacks to generate input payloads for vulnerability testing. To streamline vulnerability validation, \tool leverages large language models (LLMs) to generate targeted exploit code based on the identified vulnerabilities and generated input payloads. The LLMs serve as a tool to reduce the engineering effort required for writing test applications, demonstrating the potential of combining static analysis with LLMs to enhance the efficiency of exploit generation and vulnerability validation. Evaluation on Android 12 and 14 systems from Samsung and OnePlus demonstrates \tool's effectiveness, uncovering 51 previously unknown vulnerabilities among 217 apps with only 2 false positives. 
These results underscore the importance of combining program and access control policy analysis for accurate vulnerability detection and highlight the promising direction of integrating LLMs for automated exploit generation, providing a comprehensive approach to enhancing the security of Android systems against filesystem vulnerabilities.
\end{abstract}


\section{Introduction}
\label{sec:intro}



Mobile system platforms, such as Android, provide an operating environment for the deployment of apps to provide a wide range of functionality for the mobile device.  Mobile apps, written either by the Android OEMs or third parties, rely on the filesystem heavily to manage app resources (e.g., photos) and in some cases to share resources among multiple apps to leverage their combined functionality (e.g., take and edit photos).  


However, researchers have long known that shared access to filesystems among mutually untrusting subjects presents threats that programmers often fail to defend.  For example, researchers identified time-of-check-to-time-of-use (TOCTTOU) attacks on filesystems nearly 50 years ago~\cite{mcphee74}, which allow an adversary to trick a victim program into accessing resources that are unauthorized to the attacker.  A variety of other attack vectors have been identified, including path traversal (e.g., use unsanitized pathnames), link traversal (e.g., following untrusted symbolic links), and file squatting (e.g., using files created by an untrusted party).  For example, a recent Android vulnerability (CVE-2022-36580, see Section~\ref{sec:motivate}) is caused because programs fail to filter maliciously crafted input used to build file pathnames, causing the program to access files chosen by attackers.  


Researchers have proposed a variety of defenses to prevent filesystem attacks and methods to detect filesystem attacks, but unprotected vulnerabilities still remain common.  Several defenses were design to prevent TOCTTOU attacks~\cite{dean-hu,raceguard,openwall,payer12vee,yee03usenix,chapin_tocttou,tsafir,chari10ndss}, but researchers found that such defenses could not prevent vulnerabilities accurately (i.e., without false positives and/or false negatives) because programs lacked current information about the state of the filesystem and system lacked knowledge about programmer intent~\cite{johnson-tocttou}.  Subsequent defenses proposed methods to apply rules about program state enforced on filesystem operations to prevent vulnerabilities~\cite{process-firewall}, but such defenses have not been applied in practice.  Instead, filesystem APIs were extended (e.g., {\tt openat}) for programmers to express constraints on how filesystem operations are performed, but some unsafe behaviors are still permitted that often lead to vulnerabilities.


Researchers have also developed methods to detect filesystem vulnerabilities, using runtime analysis and access control policy analysis.  Runtime analysis techniques aim to generate test cases to detect vulnerabilities~\cite{sting12usenix,jigsaw14usenix}.  While several vulnerabilities were detected, these techniques fail to detect vulnerabilities in paths that are not executed, leaving many latent vulnerabilities unfound.  On the other hand, researchers have developed access control policy analysis~\cite{jaeger02sacmat,setools,jaeger03usenix} techniques that detect which directories may be exploited in link traversal and file squatting attacks~\cite{lee21usenix,lee2023polyscope}, but these tools cannot detect  whether programs may protect themselves from attack when using these directories.  


A key insight explored in this paper is that static analyses of programs (i.e., its construction of file pathnames) and access control policy analysis (i.e., identifying which pathnames may be attacked) can be combined to detect filesystem vulnerabilities with high accuracy.  First, static analysis and symbolic execution can be used to generate constraints on the possible file pathnames generated by the program.  Second, access control policy analysis can be used to generate constraints on the filesystem resources that attackers may use to launch exploits on filesystem vulnerabilities.  Third, modern constraint solvers~\cite{z3, cvc4} can be applied off-the-shelf to solve constraints on file pathnames to identify the file pathnames that can be exploited and the input payloads sufficient to demonstrate a filesystem vulnerability.


In this paper, we detail the design and evaluation of \tool, a tool for static detection of filesystem vulnerabilities for Android systems.  \tool combines program and access control analyses to implement the first fully static method for detecting filesystem vulnerabilities.  \tool applies Android program code (i.e., APKs) and manifest files to detect how attackers may provide malicious input to these programs and how programs handle such input to generate file pathnames (i.e., strings) used in file operations 
to access filesystem resources (i.e., files and directories).  \tool also leverages new and existing access control policies analysis based on a modified version of open-source access control policy analysis tool~\cite{lee2023polyscope} to identify constraints on the filesystem resources that may be used to attack filesystem vulnerabilities.  \tool unifies the two sets of constraints to apply off-the-shelf constraints solvers~\cite{z3} to produce payloads to drive the program to exploit filesystem vulnerabilities. Once the payload is generated using the constraint solver, it is passed to a large language model (LLM). The LLM is then tasked with writing an exploit activity that incorporates the payload, providing a practical means to verify the vulnerability. This exploit activity is designed to invoke the  the vulnerable component of the Android application and trigger the filesystem vulnerability using the generated payload. By leveraging the LLM's ability to generate code based on the provided context and payload, \tool streamlines the process of validating and demonstrating the impact of the identified vulnerabilities.


We evaluate \tool on Android version 12 and 14 systems by Samsung and OnePlus to detect filesystem vulnerabilities and generate exploit payloads.  \tool detects three kinds of filesystem vulnerabilities.  First, \tool detects {\em pathname traversal attacks}, where attackers provide malicious inputs to exploit the insufficient sanitization in programs to access private files (i.e., files that attackers are not authorized to access), avoid many false positives by assessing the Android permissions needed to provide inputs used in generating file pathnames.  Second, \tool detects file squatting and link traversal attacks, which we refer to as {\em hijacking attacks}, which enable adversaries to hijack the name resolution of a file to tamper with the program.  Third, \tool detects what we call {\em luring attacks}, which are a combination of the first two types of attacks, where attackers provide input to control the pathname used at the name resolution sink, causing the victim to access filesystem resources that have been hijacked by the adversary. In our evaluation, we test 217 programs on Samsung and OnePlus, covering more than 64\% of the pre-loaded applications.

This paper makes the following contributions:

\begin{itemize}
\item \tool is the first fully static method for detecting filesystem vulnerabilities, leverage program analysis and access control analysis to produce constraints that identify vulnerabilities and produce exploit payloads when solved. 
\item \tool detects three types of filesystem vulnerabilities automatically: pathname traversals, hijacking vulnerabilities (i.e., link traversal and file squatting), and luring traversals (i.e., pathname traversal plus link traversal).
\item \tool demonstrates the potential of combining static analysis and large language models (LLMs) for vulnerability detection and exploitation. While our current use of LLMs is preliminary, we show how providing vulnerability details and context obtained from static analysis to an LLM can help in generating customized exploit code. 
\item \tool finds 51 vulnerabilities among 217 Samsung and OnePlus apps.  These vulnerabilities include 16 path traversal, 34 hijacking vulnerabilities, and 1 luring vulnerability, with only 2 false positives.
\end{itemize}

{\bf Ethics Statement}.  We have responsibly disclosed all 51 potential filesystem vulnerabilities detected by \tool to Samsung and OnePlus. At present, four vulnerabilities have been confirmed by the vendors, including those discussed in the case studies presented in this paper. The remainder of the cases are still under investigation.  

\section{Motivation}
\label{sec:motivate}

In this section, we motivate the goals of our work by presenting examples of filesystem vulnerabilities that \tool aims to target (Section~\ref{subsec:examples}), outlining previous approaches for detecting these vulnerabilities (Section~\ref{subsec:prior}), and describing the key limitations that have prevented static detection of filesystem vulnerabilities (Section~\ref{subsec:static}).

\subsection{Examples}
\label{subsec:examples}
In this paper, we are interested in three types of filesystem vulnerabilities, described below.

\begin{figure}
\centering
\includegraphics[width=0.7\linewidth]{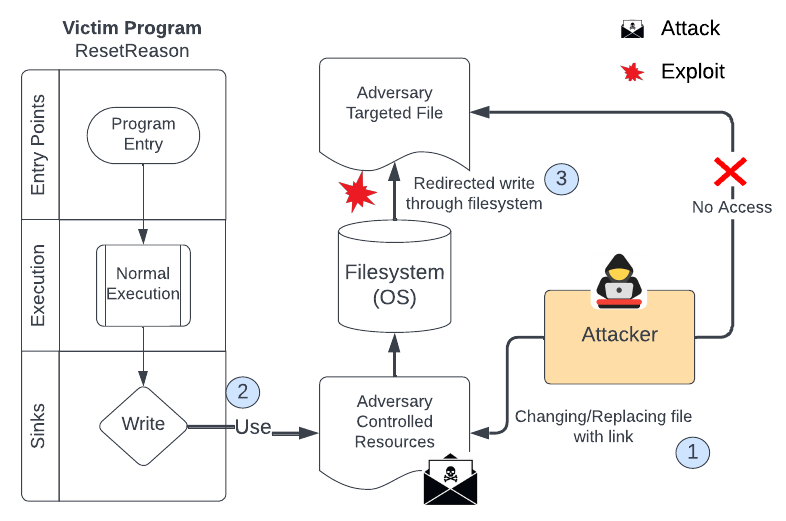}
\footnotesize
\caption{\small{\bf Filesystem Hijacking Vulnerability Example}: The file pathname used by a victim program to open a file uses an adversary-resource which redirects the victim to an adversary targeted file.}
\label{fig:file_hijacking}
\end{figure}

\paragraph{Hijacking Vulnerability}
Figure~\ref{fig:file_hijacking} illustrates the challenges in detecting vulnerabilities automatically that allow adversaries to hijack file retrieval by name using CVE-2020-13833 as an example.  In this case, the victim (ResetReason) uses a file pathname ({\tt /data/log/power\_off\_resetreason.txt}) to access a file that resides in a directory ({\tt /data/log}) to which attackers have write access (i.e., Adversary Controlled Resources in Figure~\ref{fig:file_hijacking}).  On boot, Android sends a system broadcast that causes ResetReason to open the vulnerable filepath.  Attackers can launch two types of hijacking attacks, link traversals and file squatting, to hijack the resolution of the pathname.  For example, if an attacker places a symbolic link at the vulnerable file pathname (i.e., performs a link traversal attack), ResetReason opens the target of the symbolic link instead of the expected file (i.e., Adversary Targeted File in Figure~\ref{fig:file_hijacking}).  Alternatively (not shown in the figure), attackers could create a regular file at that file pathname, enabling the attacker to control the content provided to ResetReason.  Attackers can use this vulnerability to provide attacker-controlled input to ResetReason by causing it to modify to the targeted resources.  It is difficult to detect such attacks proactively, because we lack knowledge of all the file pathnames that a program may use and/or lack knowledge of the access rights of other processes to those pathnames.  

\begin{figure}
\centering
\includegraphics[width=0.7\linewidth]{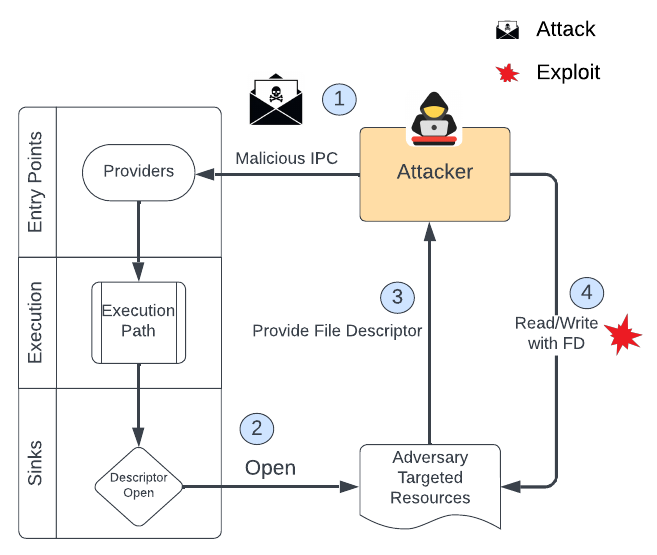}
\footnotesize
\caption{\small{\bf Path Traversal Vulnerability Example}: The victim uses malicious IPC input to construct a file pathname to a private victim file that the adversary can then read/write illicitly.}
\label{fig:directing}
\end{figure}

\paragraph{Path Traversal Vulnerability}
Figure~\ref{fig:directing} shows that attackers may direct victims to use files chosen by the attackers via a path traversal vulnerability using CVE-2021-25413~\cite{CVE-2021-25413} as an example.  In this case, the victim Contacts app has an exported component that is vulnerable. Contacts accepts external input (i.e., at Providers via IPC in Figure~\ref{fig:directing}) that is used to craft the pathname for a file operation.  If an attacker can provide such input, they can choose the file that Contacts will open, enabling attackers to obtain unauthorized access to Contact's private files, such as its internal configuration files and sensitive personal information.  Contacts itself has 54 entry points that accept external input and complex program flows that may or may not lead to file operations that use that input, making it difficult to determine whether programs may contain such vulnerabilities.

\paragraph{Luring Vulnerability}
A luring vulnerability enables an attacker to combine path traversal and hijacking vulnerabilities to choose both the file pathname to hijack and the target of the hijacking.  In some cases, programs may have some defenses to prevent the attackers from exploiting a path traversal vulnerability directly to access private files, but instead attackers can direct the victims to attacker-controlled files then hijack the file operation to target the private files or provide malicious resources.  In general, an attacker applies a path traversal vulnerability (Figure~\ref{fig:directing}) to lure the victim to a symbolic link (Figure~\ref{fig:file_hijacking}) to hijack the file operation to access an unauthorized resource.  These vulnerabilities are thus even more difficult to detect because two separate attack steps must be combined. 

\subsection{Detecting Filesystem Vulnerabilities}
\label{subsec:prior}

Researchers have long known that filesystem vulnerabilities are possible, even ones that result from time-of-check-to-time-of-use (TOCTTOU) races~\cite{mcphee74,bishop-dilger}.  Researchers have proposed a variety of defenses for TOCTTOU races~\cite{dean-hu,tsafir, chari10ndss,chapin_tocttou,raceguard}, but Cai et al.~\cite{johnson-tocttou} have shown that system-only defenses are prone to false positives when attempting to prevent all filesystem vulnerabilities because programmer intent is necessary to determine whether the access violates security.  Defenses have subsequently been developed that  
allow programmers to describe {\em how} to perform file operations (e.g., {\tt openat}).  However, some unsafe operations are still possible.  Despite the broad use of such techniques, filesystem vulnerabilities remain common, particularly for Android~\cite{MITD, CVE-2018-9587, IOACTIVE, CVE-2023-20943, CVE-2022-22292}.

Another line of research is to detect filesystem vulnerabilities in programs proactively.  These systems detect vulnerabilities in programs by generating test cases at runtime for hijacking attacks when a directory is used in name resolution that is prone to hijacking~\cite{sting12usenix} (i.e., can be written by an adversary) and by generating test cases for attacks that lack filtering of untrusted inputs used in constructing file pathnames that may lead to path traversal~\cite{jigsaw14usenix}.  A challenge is that runtime analysis does not test all filesystem uses by the programs, so while several vulnerabilities were discovered many remain latent.  In addition, runtime testing of all applications for filesystem vulnerabilities using these tools is a slow semi-automated process.


Alternatively, researchers have proposed methods to detect filesystem vulnerabilities using access control policy analysis~\cite{jaeger02sacmat,jaeger03usenix,setools}.  Access control policy analysis has been used to detect data leakage in Android systems~\cite{chen17acsac,BigMAC}.  Recently, researchers have developed methods to detect filesystem vulnerabilities with access control policy analysis~\cite{lee21usenix,lee2023polyscope}.  These works identify the filesystem resources that could be used in a hijacking attack from the access control policy (e.g., can be written by an adversary).  Once these resources are identified, one still has to determine whether these resources are actually used by the program and whether the program has defense to prevent vulnerable use.  Access control policy analysis techniques also use runtime analysis to detect vulnerabilities, with the same limitations described above for runtime analysis in general.

\subsection{Detecting Vulnerabilities  Statically and Exploit Generation}
\label{subsec:static}


To detect file filesystem vulnerabilities more comprehensively in an automated fashion, we propose to utilize static analyses.
Numerous tools and techniques have been proposed to use static analysis for identifying security vulnerabilities and malicious behaviors in programs. For Android apps, these approaches build upon general static analysis frameworks and incorporate Android-specific semantics to detect issues such as privacy leaks~\cite{flowdroid}, component hijacking~\cite{chex}, and inter-component communication vulnerabilities~\cite{intent-fuzzer, intentfuzzer}.  For example, the Chex system~\cite{chex}
uses
static analysis to evaluate components for vulnerabilities by looking for illicit data flows that lack
security checks.  However, these techniques focus on vulnerabilities due to communications among Android components rather than from components' use of the Android filesystem.  

Whether a malicious input could actually cause a filesystem vulnerability depends on the constraints that the program enforces when constructing and using pathnames in file operations.  Researchers have found that using symbolic execution enables reasoning about program constraints accurately~\cite{ma2011directed, klee, angr}.  
For example, researchers have applied symbolic execution to drive programs to explore events of interest~\cite{symJPF}, which can direct the program to determine the inputs sufficient to cause  program executions that can reach those events.  However, to detect filesystem vulnerabilities, we also need to account for the environment in which the analysis is performed, as access control policies further constrain the conditions where cause a filesystem vulnerability is possible. 

Finally, a few researchers have developed analyses that consider
constraints from the environment in addition to program constraints in
detecting vulnerabilities~\cite{siesta07atc}.  
The system that is the closest to ours is Harehunter~\cite{aafer15ccs}, which identifies cases where a malicious app may be able to receive inter-component communications (ICCs) without being vetted sufficiently to be a legitimate receiver of such communications.  Harehunter compares Android attributes declared by apps (i.e., in their manifests) to find when untrusted apps may receive ICCs, and then analyzes programs to detect whether one of the necessary security checks guards access to ICCs.  While Harehunter considers both environmental and program operations, it does not consider these constraints {\em together}, which is necessary to validate filesystem vulnerabilities. In addition, Harehunter does not generate PoC inputs
automatically to validate vulnerabilities like \tool does.

\section{Threat Model}
\label{sec:threat}
\begin{figure}[t]
\centering
\includegraphics[width=0.6\linewidth]{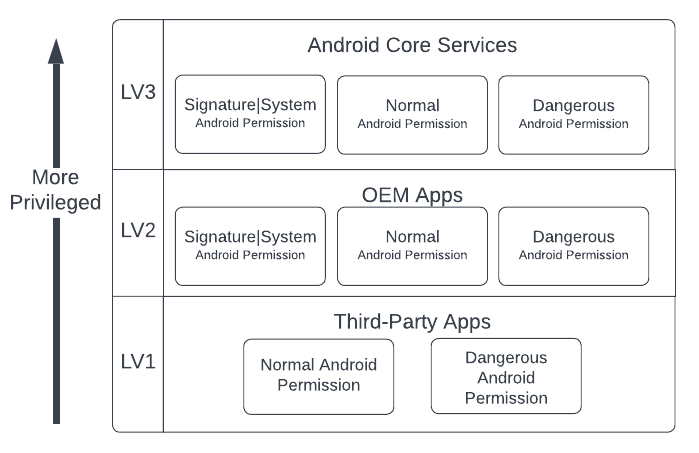}
\caption{\small{\textbf{Threat Model}: Levels of privilege and corresponding  Android permissions that each level can obtain.}}
\label{fig:threat}
\end{figure}

This work focuses on detecting filesystem vulnerabilities within OEM apps and Android core services. Android systems organize apps in terms of privilege levels~\cite{PPRIV_LEVEL}, where we consider three privilege levels as shown in Figure~\ref{fig:threat}.  Third-party apps are at privilege level 1 (LV1), OEM apps are at privilege level 2 (LV2), and Android core services are at privilege level 3 (LV3).  Further, 
Android permission system groups permissions into three categories: (1) normal, (2) dangerous, and (3) system/signature permissions.  Normal permissions are available to all apps by default.  

Our threat model makes two key assumptions about the Android permissions assigned to apps at distinct privilege levels:
\begin{enumerate}
    \item Third-party apps (LV1) are expected to declare as many normal and dangerous permissions as possible, assuming all will be granted by some user (i.e., assume a worst-case attacker). However, they cannot obtain system/signature permissions.
    \item Each OEM app (LV2) and Android core service (LV3) has a fixed set of permissions declared in individual manifest files.
\end{enumerate}

LV2 and LV3 apps often use Android permissions to control access to their exported components (i.e., control inputs to those apps via their components), which may be at the system/signature level (i.e., not available to LV1 apps). This ensures that only properly credentialed apps can interact with sensitive components.  In this paper, we model control of exported components (i.e., application entry points) as {\em entry constraints} as described in Section~\ref{sec:design}.

In this paper, we consider two main threat scenarios. First, attacks from lower-privileged apps to higher-privileged apps. This includes attacks from LV1 apps to LV2 and LV3 apps, as well as attacks from LV2 apps to LV3 apps. his is the default Android threat model, and successful attacks on LV2 or LV3 apps can result in significant damage~\cite{lee21ieeesp}, as some of these apps have file permissions that extend beyond the application sandbox. Second, we consider attacks among LV1 apps, as some security-critical apps, such as banking applications, run at this privilege level.

\section{Filesystem Vulnerability Detection Problem}
\label{sec:Overview}

\begin{figure}
\centering
\includegraphics[width=0.9\linewidth]{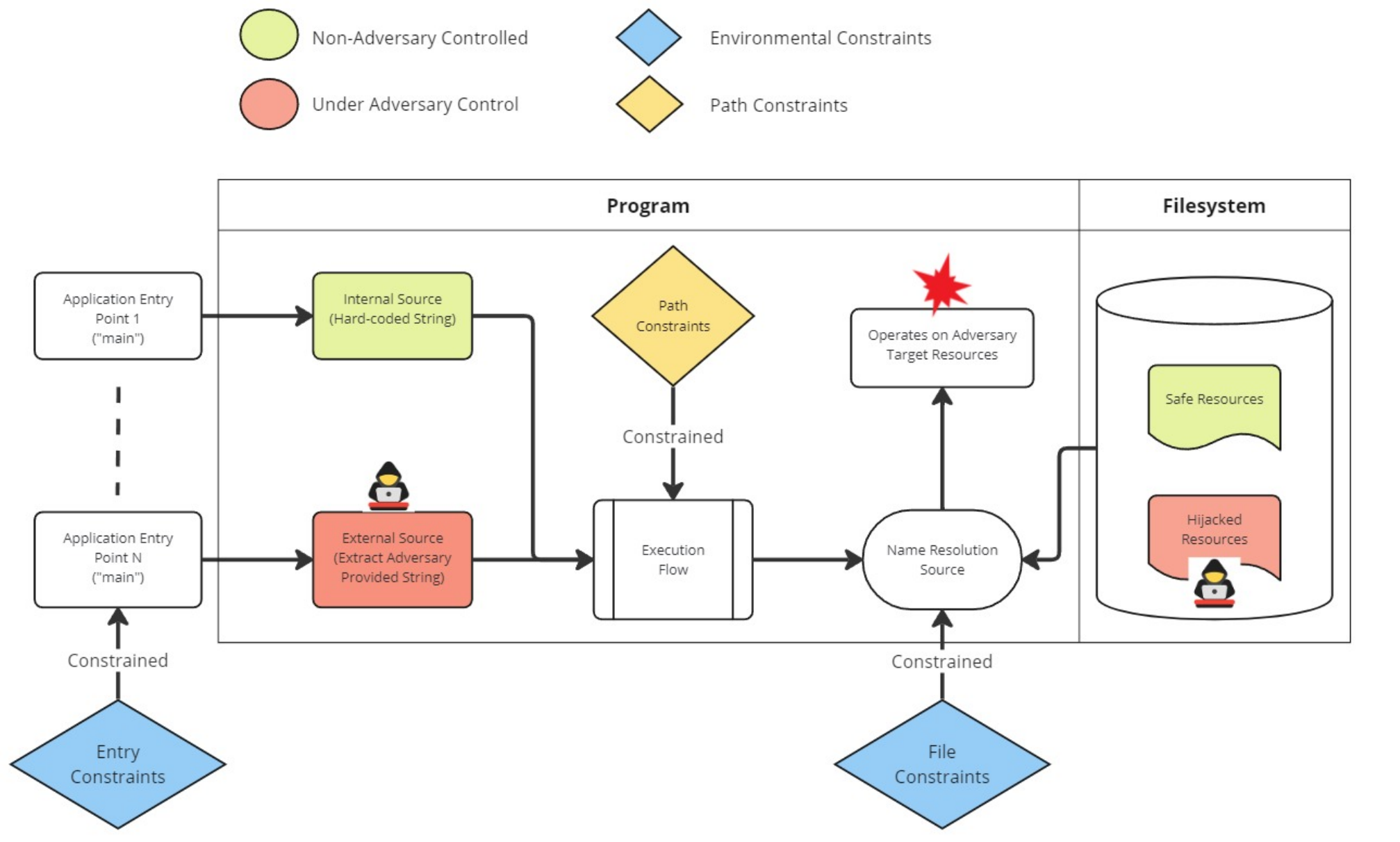}
\caption{{\bf Filesystem Vulnerability Detection Problem}: 
Attackers may attack at {\em external sources} and {\em name resolution} of file operations to direct the program to adversary targeted resources.  Vulnerabilities are possible if the Android system's {\em environmental constraints} and the program's {\em path constraints} allow the generation of vulnerable pathnames at file operations.  
}
\label{fig:design_overview}
\end{figure}

To identify filesystem vulnerabilities effectively, we first need to characterize the key conditions that must be satisfied to make the attacks possible. Figure \ref{fig:design_overview} illustrates how threats may impact the way that potential victim programs in Android systems (i.e., OEM apps and Android core services) perform filesystem operations.  Unlike traditional programs with a single "main" entry point, Android programs have multiple entry points defined as components in the AndroidManifest file, such as Activities, Services, Broadcast Receivers, and Content Providers.  These components allow the program to be invoked in multiple ways.  From these entry points, programs may obtain inputs from sources (i.e., internal and/or external) to construct pathnames.  These pathnames are then used by filesystem operation sinks to resolve those pathnames (i.e., via name resolution) to retrieve filesystem resources.

\paragraph{\bf Threats to File Access}
To access (e.g., open) a file, a program uses input from {\em sources} to construct pathnames that are applied at file operation {\em sinks}. We distinguish two types of sources of potentially vulnerable pathnames: internal and external.  An {\em internal source} is defined by the program, typically a hard-coded string.  An {\em external source} (red in Figure~\ref{fig:design_overview}) receives input from outside the program, such as from an entry point that may receive interprocess communications (IPCs) from third-party apps.  Android provides functions for extracting such inputs (e.g., via functions like {\em GetStringExtra} for IPCs). External sources that are accessible to potential adversaries are part the program's attack surface, whereas internal sources are not.  

In this work, the sinks are file operations that perform {\em name resolution}.  File operations that perform name resolution use a pathname provided by the program to retrieve a filesystem resource (e.g., obtain a file descriptor for accessing the file).  The filesystem converts file pathnames provided by the program into filesystem resources (i.e., files and directories) by resolving the pathname element-by-element (e.g., one directory at a time) using the filesystem's namespace (i.e., the mappings of names to filesystem resources).  Since attackers may control parts of the filesystem used in name resolution, these file operations or {\em name resolution sinks} shown in Figure~\ref{fig:design_overview} are also part of the program's attack surface.  

\paragraph{\bf Constraints on File Access}
Defenders can limit the ability of adversaries to exploit vulnerabilities from these two attack surfaces by constraining the input, construction, and resolution of pathnames in filesystem operations. These constraints can be categorized into two main types: environmental constraints and path constraints.

{\em Environmental constraints} describe restrictions defined outside of the program. These constraints can be further divided into two categories: entry constraints and file constraints. {\em Entry constraints} restrict which subjects may be able to provide input to external sources. For example, OEMs can define custom Android permissions to restrict access to components, e.g., preventing third-party applications from sending IPCs thereby protecting the program from input to external sources from third-party apps. In addition, Android enforces multiple access control mechanisms that may prevent attacks on name resolution. We describe such constraints as {\em file constraints}, which limit both attackers' and victim's access to filesystem resources. Researchers have previously found that to perform a hijacking attacks, adversaries need write access to a directory used by a victim in name resolution~\cite{sting12usenix}.

{\em Path constraints} describe the restrictions defined by the conditional branches within the program.   These constraints, as is typical for symbolic execution engines~\cite{angr, klee, jbse}, restrict the values of program variables when that conditional branch is taken, which may prevent the creation of exploitable pathnames used at sinks.  For example, a conditional may require that all pathnames have a particular prefix to limit file operations to a subset of the filesystem's namespace and/or are free of special characters that may enable escaping this namespace restriction.  Such path constraints may prevent adversaries from supplying inputs that enable them to choose unauthorized files for exploiting path traversal or luring vulnerabilities. 

By considering both environmental constraints and path constraints, defenders can analyze potential vulnerabilities in file operations, taking into account the restrictions imposed by the program flows as well as environmental factors, such as the Android permission system.

\paragraph{Challenges in Detecting Filesystem Vulnerabilities}
The goal of this paper is to detect filesystem vulnerabilities to path traversal, hijacking, and luring  (see Section~\ref{subsec:examples}) accurately.  To achieve this goal, our aim is to generate payloads to supply to entry points that cause the program to generate pathnames that may be exploited in these attacks.  To our knowledge, this is the first work that proposes to develop a static analysis to detect these filesystem vulnerabilities, as there are several challenges to overcome.


First, as seen in Figure~\ref{fig:design_overview}, Android programs may have multiple entry points and a variety of file operations.  A simple taint analysis would find many of the flows from entry points through sources to sinks, but we find that only a small fraction of these flows may actually result in vulnerabilities.  Android provides a variety of defenses to prevent these attacks, including custom Android permissions to restrict the invocation of entry points, methods to sanitize input used to build file names (e.g., getCanonicalPath), and methods to prevent the use of symbolic links (e.g., filesystem configurations). However, programmers apply these defenses in ad hoc ways, which may lead to vulnerabilities.  We aim to apply such Android-specific knowledge to prune flows that cannot possibly result in vulnerabilities to reduce analysis effort without missing any vulnerabilities.

Second, programs may construct and restrict pathnames in a variety of complex ways.  We must devise methods to collect the possible path constraints that determine the pathname values 
from a combination of hard-coded and symbolic sources.  Research on constraint solvers for strings has improved significantly in recent years~\cite{string_solving}, enabling the solution of complex path constraints on strings.  However, we must be aware of the various Android defenses to generate accurate path constraints that minimize the possibility of false positives in the analysis.

Third, although we can collect and solve string constraints, we need to determine which pathnames may cause vulnerable file operations on an Android system.  Android systems have an extensive, fine-grained access control mechanism, consisting of mandatory access control (e.g., SEAndroid~\cite{seandroid}), Linux discretionary access control, and specialized Android controls, such as the Android permission system~\cite{android-permissions} and Android Scoped Storage~\cite{lee21ieeesp}.  Access control policy analysis~\cite{jaeger02sacmat,setools} ingests access control policies to compute the filesystem resources that may be used in attacks.  Researchers have explored how to use access control analysis to detect the resources that may be used in some Android filesystem attacks~\cite{lee21usenix,lee2023polyscope}, but other attacks like path traversal have not been studied yet.  In addition, prior research has not used the results of access control analysis to detect program vulnerabilities statically.

\vspace{-0.05in}
\section{\tool Design}
\label{sec:design}
In this section, we examine the key elements of \tool that facilitate filesystem vulnerability discovery.

\vspace{-0.05in}
\subsection{\tool Overview}
\label{subect:tool}

\begin{figure*}
    \centering
    \includegraphics[width=\linewidth]{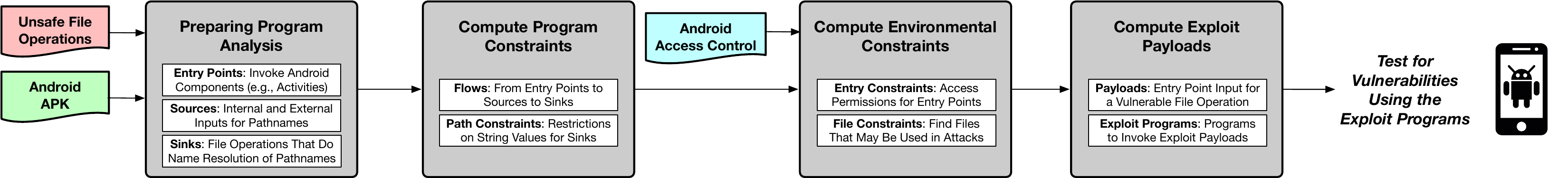} 
    \caption{{\bf \tool} processes Android APK files (green) and sensitive filesystem operations (red, defined manually) to 
    compute program (path) constraints and processes the Android access control policies (blue) and APKs (green) to compute environmental (entry point and file) constraints.  Exploit programs are generated from solutions to this combined constraint set to test for each vulnerability dynamically.}
    \label{fig:overview}
\end{figure*}

\tool operates as shown in Figure~\ref{fig:overview}.  \tool first prepares the analysis steps by extracting an Android program's entry points and sources from its APK file and identifies sinks from a list of sensitive file operations specified manually.  Second, \tool computes a program-dependence graph (PDG) representation to perform the program analysis, which generates path constraints that describe restrictions on the pathnames that may be produced when the program is executed from an entry point.  Third, \tool performs analyses of the Android environment (e.g., access control policy analysis) to compute the entry point and file constraints that dictate which entry points may be invoked by attackers and which filesystem resources may be under attacker control, respectively.  Fourth, \tool solves the combined constraint set to generate solutions from which exploit programs are produced.  Solutions are generated automatically, but the exploit programs are generated semi-automatically with the aid of Large-Language Models (LLMs).


After introducing how the analysis inputs (i.e., entry points, sources, and sinks) are prepared, we describe the \tool analyses (program and environment) in terms of how it generates exploits for the three classes of filesystem vulnerabilities in this paper, path traversal, hijacking, and luring.  We then discuss how we enable \tool to generate exploit programs from the resultant constraint sets.

\vspace{-0.05in}
\subsection{Analysis Pre-Processing}
\label{subsec:pre}

\tool performs a variety of pre-processing to collect inputs for use in vulnerability detection, including: (1) identifying name resolution sinks and (2) finding the program entry points and external sources of the  program inputs.  

\paragraph{Name Resolution Sinks} In the process of identifying critical sinks for our analysis, we rely on a recent empirical study of filesystem vulnerabilities, which provides insight on their patterns~\cite{yu2024file}. This study categorized file operations subject to attack into six distinct types. The proportion of each operation type, as derived from real-world data, is detailed in Table~\ref{table:ops}.

\begin{center}
\begin{threeparttable}
\caption{Classes of File Operations Prone to Vulnerabilities}
\label{table:ops}
\newcolumntype{Y}{>{\centering\arraybackslash}X}
\vspace{-0.1in}
\begin{tabularx}{\linewidth}{|Y|Y|}
\hline
\textbf{Operation Type} & \textbf{Percentage} \\
\hline
Process Creation & 28.4\% \\
Image Loading & 45.1\% \\
Moving & 1.1\% \\
Reading & 7.1\% \\
Creating & 8.2\% \\
Deleting & 10.1\% \\
\hline
\end{tabularx}
\end{threeparttable}
\end{center}

Given the insights from the empirical study, our selection of sinks focuses on the five classes of operations that are susceptible to attacks on Android systems: moving, creating, deleting, reading, and image loading. This strategy intentionally omits process creation due to its relative infrequency in Android environments, thus tailoring our approach to the platform-specific characteristics. We incorporate a comprehensive range of Java file operation mechanisms, prominently including classes from \texttt{java.io.File}, to capture the diverse modalities of file interaction prevalent in Java APIs. Additionally, our selection extends to encompass frequently utilized Android API methods, such as \texttt{ParcelFileDescriptor}.

\begin{figure}
    \centering
\includegraphics[width=\linewidth]{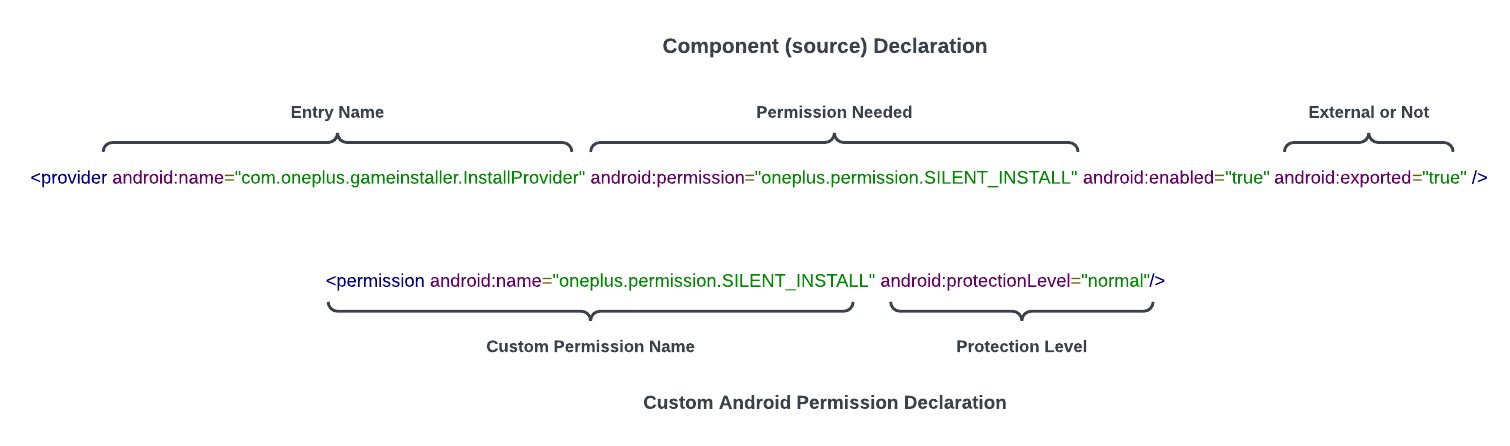} 
    \caption{Android Manifest Definition for an Exported Entry Point and Its Android  Permission}
    \label{fig:source_con}
    \vspace{-0.2in}
\end{figure}

\paragraph{Program Entry Points} 
Identifying the program entry points that lead to external sources is fundamental for detecting path traversal and luring vulnerabilities in Android applications. Program entry points refer to application components declared in the AndroidManifest file, which accept input from outside the program, such as IPC mechanisms, and are accessible to potential adversaries. Examples of these components include Activities, Services, Broadcast Receivers, and Content Providers. It is important to note that this work does not consider dynamically registered components, and we leave them as future work. We extract these entry points 
from the application's AndroidManifest file. 
Figure~\ref{fig:source_con} shows an example of entry point declaration, where the first line defines the entry point's name, whether it is external (i.e., "exported" set to true), and the Android permission needed to access the entry point.  We will return to this example when we examine the use of entry constraints.  
The intents retrieved from an exported entry point with the function {\tt Context.getIntent()} enable the program to gather the inputs provided to that entry point.  Any string retrieved from that function is deemed an external source.  Any use of {\tt Context.getIntent()} in the control flow from an attacker-accessible entry point is an external source.

\subsection{Detecting Path Traversal Vulnerabilities}
\label{subsec:design_direct}

In path traversal vulnerabilities, an attacker provides input at an external source that can direct the program to use a resource to which the attacker normally lacks access.  To detect a path traversal vulnerability, we need to: (1) determine which entry points may enable attackers to provide input at external sources; (2) collect the constraints on the pathnames that can be constructed at any reachable file operations (sinks); and (3) compute a solution for the input payload that enables access to a private program resource (i.e., a resource normally inaccessible to the attacker) at a name resolution sink.  Past work has detected path traversal attacks using dynamic analysis~\cite{jigsaw14usenix}, but this work detects vulnerabilities as the lack of {\em any} filtering code that sanitizes the input.  Instead, \tool computes the exact filtering the program performs and determines whether those filters can still circumvented to access private files.

\paragraph{Find Attacker Entry Points} The first task is to determine which entry points may be accessible to attackers to provide external sources.  Although many entry points may be exported, only a few may be accessible to attackers.  Past work does not consider methods for computing the accessibility of entry points statically.
Returning to Figure~\ref{fig:source_con}, which shows the entry point definition in an APK's manifest, the second line shows the class of the permission, which is declared as "normal" here (i.e., out of normal, dangerous, and system/signature, see Section~\ref{sec:threat}).   Using this information and the Android permissions declared for each application (of LV1 and LV2), \tool computes the set of subjects that have the Android permissions necessary to access each entry point, which forms the entry constraint.  We note that this second line of Figure~\ref{fig:source_con} may be defined in any application's manifest file, so we need to record the category (i.e., protection level) of every Android permission first and then determine the entry constraints.
The entry constraints may eliminate the need to perform analysis for some entry points and their sources, focusing only on the ones that are accessible to attackers.  For example, out of 30 entry points in the Samsung ServiceMode app 
only two are accessible to adversaries.

\paragraph{Compute Path Constraints} The second task is to collect the constraints on the pathnames that can be constructed at any file operation sinks reachable from these entry points.  In general, the path constraints collected by symbolic execution can be used to compute the possible pathnames that could be generated at any particular reachable sink.  However, symbolic execution has known scalability issues, so we first compute the control flow paths from entry points to external sources to sinks statically, and then apply symbolic execution.  To determine whether an external source is used as a pathname by a file operation sink, we perform a taint analysis over a Program Dependence Graph~\cite{liu17ccs} (PDG).  That is, using the PDG, we determine whether there is a control flow from the entry point to the external source to the sink that includes a data flow from an external source to a (pathname) sink.  We then direct symbolic along that control flow  to collect all the constraints on inputs enforced in that control flow.

\paragraph{Compute File Constraints} The third task is to compute the values of the inputs necessary to enable the program to access a file that enables a path traversal attack.  A path traversal attack enables an attacker to direct the victim to access a file that would normally be inaccessible to the attacker, which is a form of a {\em confused deputy attack}~\cite{confused-deputy}.  Past dynamic testing methods for such attacks chose a single example file (e.g., {\tt /etc/shadow}) as an exemplar of an inaccessible file, but this approach may miss attacks on other private files.  In general, we need to know the files that are accessible to the program, but inaccessible to its attackers who can use the entry point, i.e., {\em the set difference between the files accessible to the victim and each attacker}.  The file constraint is the union of these set differences, as there is a vulnerability if any attacker can trick the victim into accessing a file that is inaccessible to the attacker. 

Researchers have developed access control policy analysis methods to compute the resources accessible to individual subjects~\cite{jaeger02sacmat,setools}.  We utilize a recent access control policy analysis tool for Android, called PolyScope~\cite{lee21usenix,lee2023polyscope}.  PolyScope computes the authorized file operations to a program that may be modified by an attacker.  PolyScope further accounts for some attacker changes in policy and user expansion of program permissions (e.g., via Android permissions),
as DAC models permission assignments are not limited by design~\cite{take-grant,tam} or policy constraints~\cite{tidswell00ccs}.
However, PolyScope is only designed to compute the files that are accessible to both the victim and the attacker, so we adapted PolyScope for \tool to compute the set difference. To compute a solution for the input payload, we combine the path and file constraints and apply a constraint solver to generate a solution, if one exists.  We used the Z3 solver for this task.




\subsection{Detecting Hijacking Vulnerabilities}
\label{subsec:design_hijack}

In a hijacking vulnerability, internal sources define the input to create the file pathnames used at file operation sinks, but if these file pathnames include directories that an attacker can modify, then an attacker may hijack the name resolution of these file operations.  There are three tasks for detecting hijacking vulnerabilities efficiently: 
(1) determine the hard-coded strings that may used as internal sources; (2) determine whether a hard-coded pathname includes a directory that is modifiable by an attacker; and (3) generate path constraints from the entry point to the sink.  We also need to determine whether the entry point that uses an internal source is accessible to an attacker, but this is determined using entry constraints in the same way as for path traversal attacks. 

\paragraph{Compute Internal Sources} First, we must identify the hard-coded strings that may be used as internal sources of pathnames in file operation sinks.  As there may be many hard-coded strings in programs and not all may be used in file operations, we propose to use a reverse data-flow analysis on the program's PDG to find the relevant strings used as internal sources, starting from the sink's pathname argument.  If the reverse data-flow analysis finds an external source as input, then \tool assumes this flow is assessed for path traversal vulnerabilities, as described above.  

\paragraph{Compute Attackable Internal Sources}  Second, to find hijacking vulnerabilities, we must determine whether any attacker with access to the entry point has write access to an element (i.e., directory) of the internal source string.  We once again utilize PolyScope~\cite{lee21usenix,lee2023polyscope} to {\em find the directories in the hard-coded pathname to which attackers have write access}.  This is a task PolyScope performs already. 

However, to reduce the analysis effort, we prune flows where an internal source cannot possibly be exploited. This is only possible if the internal source includes the name of a directory to which an attacker has write access. 
Thus, we check the pathname provided by this internal source directly against the file constraints to prune cases that do not use adversary-controlled directories.  
By applying this approach, we can prune sinks from the analysis do not use any adversary-controlled directories before we spend the effort generating path constraints.  

\paragraph{Compute Path Constraints} It is important to emphasize that the goal is to generate inputs at an entry point to drive the program execution to use a vulnerable internal source at a sink.  To do this, \tool starts the symbolic execution to generate path constraints from entry points rather than from internal sources.  This approach ensures that all the relevant conditions along a valid execution path are captured. For internal sources that can be reached from multiple program entry points, including those involving inter-component communication (ICC), we perform symbolic execution from all these entry points.  We then combine ("and") the path constraints over the file pathname argument in the sink with the file constraints for each entry point independently, to determine whether the file pathname generated from internal sources from any entry point would use filesystem resources under adversary control, resulting in a hijacking vulnerability.

\subsection{Detecting Luring Vulnerabilities}
Luring vulnerabilities combine aspects of both path traversal and hijacking vulnerabilities to deceive the victim program into accessing private files.  On the one hand, the need to direct the program to use a file pathname of the attackers' choosing, much like a path traversal vulnerability.  However, in this case, the program is not susceptible to the simpler path traversal vulnerability.  Instead, in a luring vulnerability, attackers direct the victim to attacker-controlled files to redirect (e.g., via a symbolic link) the victim program's name resolution to a private file.  
Thus, to detect luring vulnerabilities, \tool performs the same analysis as for path traversal vulnerabilities, as described in Section~\ref{subsec:design_direct}, but instead of directing the victim to a private file, a luring attack must first lure the victim to an adversary-controlled directory as in a hijacking vulnerability.  Thus, the only difference between the analysis to detect a luring vulnerability rather than a path traversal vulnerability is the use of file constraints from the hijacking vulnerability detection to identify directories to lure the victim to directories that can be hijacked.  

\subsection{Exploit Payload Generation}
\label{subsec:LLM}

The ultimate goal of this research is to produce an exploit program that can be used to test for and demonstrate the vulnerability. We use a constraint solver~\cite{z3} to produce an exploit payload, when possible given the combination of file and path constraints computed as described above. It is important to note that Android payloads are not simple command line inputs; instead, we need to construct Android IPC in the correct format, which involves non-trivial engineering work.

Large language models (LLMs) have shown promise in generating code snippets that help with testing and understanding natural language instructions. However, they often struggle with processing and generating large, complex codebase~\cite{ozkaya,wang2024llm}, which can hinder their effectiveness in creating full-fledged exploits directly from the codebase, especially when dealing with Android applications. LLMs also face challenges in identifying vulnerabilities related to inter-component communication (ICC), a common attack vector in Android systems, as it requires a deep understanding of the application's structure and component interactions.

To address these limitations, we enhance exploit generation by leveraging synergy between \tool's analysis and LLMs to create precise and effective Android exploit programs. Our method begins by utilizing \tool's comprehensive static analysis to gather critical information about the target application, including the vulnerable component's structure, the exact method invocation path to the file operation sink, and the payload produced by solver that would drive to program to desinated sink.

We use the information gathered from static analysis, including the method invocation path, decompiled code, and Android Manifest details, to construct a comprehensive prompt for the LLM. The prompt explicitly outlines the required intent structure (action, flags, and categories) and any necessary extra data fields (extras, URIs, or bundles). By incorporating this contextual information, the LLM can generate exploit code that accurately navigates the application's control flow to reach the vulnerable sink. Guided by the static analysis, the LLM provides usable code for all positive cases. Although some minor bugs may need to be fixed in certain instances, these typically only require adjustments to a few lines of code.

With our approach, the LLM generates complete, runnable Java code for an exploit activity, leveraging the tight integration of static analysis results to produce highly targeted exploits tailored to specific vulnerabilities. This potentially increases the effectiveness and precision of the exploits while significantly reducing the manual effort required for their creation.

\section{Implementation}
\tool is implemented on top of Soot~\cite{soot}, FlowDroid~\cite{flowdroid}, and Intellidroid~\cite{intellidroid} using Java and Python. In terms of choice of LLM, we leverage Google's Gemini~\cite{gemini}. We leverage Soot to generate an intermediate representation (IR) of the Android application, which serves as the foundation for our directed symbolic execution. Soot's extensibility allows us to override methods used to interpret the IR, enabling us to customize the symbolic execution process according to our needs.
One of the key features of our implementation is the ability to intercept specific function invocations. We take advantage of this capability to emulate commonly used Android method calls mentioned in Section~\ref{sec:design}, such as \texttt{getExternalFileDir()} and \texttt{getFilesDir()}. By intercepting these method calls and resolving them to concrete values whenever possible, we enhance the precision of our symbolic execution and improve the accuracy of the collected path constraints.
To improve the entry-point discovery mechanism, we extend FlowDroid's existing functionality to consider entry constraints. By incorporating entry constraints into the entry-point analysis, we can distinguish between internal and external sources, providing a more comprehensive understanding of the application's attack surface. This extension enables us to identify and prioritize external sources that are accessible to potential adversaries, focusing our analysis efforts on the most critical entry points.
The combination of Soot, FlowDroid, and Intellidroid provides a robust foundation for \tool, greatly reducing our development workload. By leveraging the strengths of these existing tools and their modular designs, we were able to implement \tool with ~2,000 lines of additional code. This efficiency is a testament to the careful planning and consideration given to future extensibility by the developers of Soot, FlowDroid, and Intellidroid. 

\paragraph{Analysis initialization} Before starting the program analysis, we perform two essential pre-computation steps: entry constraint analysis and file constraint analysis. We identify program entry points that lead to external sources, and extract the corresponding entry constraints. Additional taint analysis will be performed on these external sources to verify whether adversary input reach targeted sinks. File constraints, collected using PolyScope, identifies per-app dangerous resources. By incorporating file constraints, we can focus on analyzing execution paths that perform name resolution on dangerous filesystem resources, reducing the number of paths requiring symbolic execution.

\paragraph{Modeling Program Defenses}
One key function we emulate is \texttt{getCanonicalPath()}, which plays a critical role in mitigating path traversal attacks. Our study of the most recent 15 path traversal vulnerability CVEs~\cite{CVE_directory_traversal} reveals that all patches involve converting adversary-controlled input into a canonical format, effectively neutralizing ".." or "." sequences that could be exploited to navigate to unauthorized directories. In our analysis, when \texttt{getCanonicalPath()} is invoked on adversary-controlled input, we mark the corresponding path as safe from directing vulnerabilities, as this function eliminates many for path traversal attacks.  However, this is a complex function and may have flaws, which we will explore further in the future.

\paragraph{Handling Android Idioms}
Some Android idioms present challenges for generating path constraints.  We describe how we address the two main challenges.  First, Android programs use a variety of Android APIs to generate pathnames.  For example, the function {\tt getExternalStorageDirectory} returns the concrete value \texttt{"/storage/emulated/0"}.  We produce summaries of these functions to generate the expected outputs.  Second, some Android APIs convert specific pathnames to indices to guide the program execution more efficiently when those pathnames are used.  However, \tool wants to reason about path constraints in terms the pathnames rather than integers.  As a result, we apply Soot to analyze the class's static initializers to map the integer values to the expected pathnames when generating constraints.

\paragraph{Exploit Generation and Testing}
To generate inputs that drive the victim program to potential vulnerabilities, we solve the collected constraints using the off-the-shelf SMT solver Z3~\cite{z3}. The solver provides the necessary input values to construct appropriate test cases. Together with data extracted from the apk, including the manifest declaration of the entry point and decompiled component code with JADX~\cite{jadx}, we instruct the LLM to generate a test activity that would trigger the vulnerable code in the victim program. We then perform testing on the device using the generated exploit code. However, some victim components may be protected by system or signature-level permissions, which third-party apps cannot directly interact with. In these cases, we use a rooted device with Magisk to escalate privileges and perform testing on these protected components as the root user. This approach allows us to comprehensively test all identified paths and components, regardless of their privilege levels. It is important to note that the additional permissions are necessary to mimic attackers at LV2 with signature permissions. For testing exploits from LV1 attackers targeting unprotected components, these elevated privileges are not required.

\section{Evaluation}
\label{sec:eval}

We conducted an analysis using our tool on OEM applications from two different vendors: Samsung and OnePlus. The objective of our evaluation was to determine the prevalence of filesystem vulnerabilities in pre-installed OEM applications. We specifically selected applications that operate with elevated privileges—those that possess permissions inaccessible to attackers. Table~\ref{tab:overall} presents the number of applications analyzed from each vendor. It is important to note that some applications do not have sensitive sinks (e.g., no file operations) and were excluded from the average counts. Additionally, there were applications that caused our tool to terminate during the PDG construction phase, likely due to extensive obfuscation.

Based on the average source and sink counts detailed in Table~\ref{tab:overall}, OEM applications show a moderate level of complexity and file usage. While the number of source-to-sink flows is not exceptionally high, it's noteworthy that these applications have multiple entry points and utilize file operations regularly. Given that these are pre-installed applications with elevated privileges, it remains important to assess the security of these file operations.

\begin{table}[t]
\centering
\resizebox{0.9\columnwidth}{!}{%
\footnotesize
\begin{threeparttable}
\caption{General App Statistics}
\label{tab:overall}
\begin{tabular}{lccc}
\toprule
\textbf{Devices} & \textbf{Samsung 12} & \textbf{Samsung 14} & \textbf{OnePlus}\\
\midrule
Total Apps & 72 & 82 & 63 \\
Apps without Sinks (no file operations) & 5 & 15 & 19 \\
Apps Causing Analysis Failure & 7 & 10 & 9 \\ \hline
Total Apps Tested & 60 & 57 & 35 \\ \hline
Average Entry Count & 26  & 31  & 19  \\ \hline
Average Source Count $^1$& 37/29/8 & 41/34/7 & 44/38/6 \\
Average Attackable Source Count$^1$ & 0.45/0.28/0.17 & 0.24/0.18/0.06 & 0.6/0.57/0.03 \\
Average Sink Count & 31 & 38 & 47 \\
Average Source-to-Sink Flow Count & 33 & 36 & 43 \\
\bottomrule
\end{tabular}
\begin{tablenotes}
\scriptsize
\item $^1$ Total/Internal/External
\end{tablenotes}
\end{threeparttable}
}
\end{table}

\subsection{Vulnerability Overview}
Table~\ref{tab:vuln_overview} presents the results of our analysis on OEM applications from Samsung and OnePlus. 

\begin{table}[htbp]
\centering
\resizebox{0.9\columnwidth}{!}{%
\footnotesize
\begin{threeparttable}
\caption{Vulnerability Detection Result}
\label{tab:vuln_overview}
\begin{tabular}{lccc}
\toprule
\textbf{Devices} & \textbf{Samsung 12} & \textbf{Samsung 14}  & \textbf{OnePlus 12} \\
\midrule
Total Apps Tested & 60 & 57 & 35 \\ \hline 
Hijacking Positives & 13& 7 & 16 \\
Hijacking True Positives & 11& 7 & 16 \\ \hline
Path Traversal Positives$^1$ & 13 & 2 & 1 \\
True Positive Path Traversal Vulns & 13& 2 & 1 \\ \hline
Luring Positives$^1$ & 0& 0 & 1 \\
Luring True Positives   & 0 & 0 & 1 \\
\bottomrule
\end{tabular}
\begin{tablenotes}
\scriptsize
\item $^1$ count in number of control flow paths
\end{tablenotes}
\end{threeparttable}
}
\end{table}

\paragraph{Validating True Positives}
\tool produces input payloads for vulnerability (positive) that it identifies.   Such payloads are designed to cause victim programs to use adversary-chosen resources for each of the three classes of vulnerabilities.  We state that when a victim executes the payload and uses (e.g., {\tt open}s) an adversary-chosen resource, then it has a  filesystem vulnerability (i.e., has a true positive).  
Thus, to confirm true positives, we manually validate each case by running the LLM-generated driver activity (i.e., payload) on-device and observing its behavior. A case is classified as a true positive if the victim program opens the adversary-chosen filesystem resource.
We discuss examining how such resources are then {\em used} by the victim program in Section~\ref{subsec:accuracy}. 

For each class of vulnerabilities, validation differs slightly.  For hijacking vulnerabilities, we expect that a particular adversary-controlled file or link is used in the {\tt open} operation.  For path traversal vulnerabilities, true positives occur when the payload drives the victim program perform to a file operation sink to access an unauthorized (to the attacker) resource. 
Luring vulnerabilities are similar to path traversal vulnerabilities, but with an additional requirement: the victim program must use an adversary-controlled symbolic link to direct the victim to an unauthorized (to the attacker) resource.  

\subsection{Effectiveness of Environmental Constraints}
\label{subsec:file_constraint_triage}
Environmental constraints demonstrate significant triage power in our analysis, drastically reducing the attack surface. File constraints eliminate flows with sinks using safe file paths, while entry constraints filter out paths with inaccessible sources. As shown in Table~\ref{tab:overall}, the average number of attackable sources per app is significantly reduced compared to the total sources. For instance, in Samsung 12 devices, only 0.45 out of 37 average sources per app are identified as attackable after applying constraints.
Table~\ref{tab:eff_constraint} illustrates the impact on source-to-sink flows, showing a reduction of over 96\% across all devices. This dramatic decrease in analyzable paths allows security analysts to focus on a smaller, more relevant set of potentially vulnerable paths, significantly enhancing the efficiency and effectiveness of the vulnerability detection process.
\begin{table}[htbp]
\centering
\resizebox{0.9\columnwidth}{!}{%
\begin{threeparttable}
\caption{Flow Before Environmental Constraints vs. After}
\label{tab:eff_constraint}
\renewcommand{\arraystretch}{1.2}
\setlength{\tabcolsep}{2.1pt}
\begin{tabular}{@{\hspace{0pt}}lcccc@{\hspace{0pt}}}
\toprule
\textbf{App Name} & \textbf{Samsung 12} & \textbf{Samsung 14} & \textbf{Oneplus} \\
\midrule
Source-to-Sink Flow Count$^1$ (Total Pre) & 1,980 & 2,052 & 1,505\\
Source-to-Sink Flow Count (Total Post) & 72 & 44 & 61\\
\bottomrule
\end{tabular}
\begin{tablenotes}
\scriptsize
\item $^1$ Simple PDG traversal without environmental constraints
\end{tablenotes}
\end{threeparttable}
}
\end{table}

\subsection{Case Studies}
\label{subsec:case_study}

To further demonstrate the effectiveness of our tool and the real-world impact of the vulnerabilities it can detect, we present case studies on three interesting Android applications: SecTelephony (Samsung Android 12), Wallpaper (Oneplus Android 12), and GameInstaller (Oneplus Android 12), and confirmed vulnerabilities associate with these apps.

\begin{table}[htbp]
\centering
\resizebox{0.8\columnwidth}{!}{%
\footnotesize
\begin{threeparttable}
\caption{Deep Dive into Vulnerable Apps}
\label{tab:case_study}
\renewcommand{\arraystretch}{1.2} 
\setlength{\tabcolsep}{2.1pt} 
\begin{tabular}{@{\hspace{0pt}}lcccc@{\hspace{0pt}}} 
\toprule
\textbf{App Name} & \textbf{SecTelephony} & \textbf{Wallpaper} & \textbf{GameInstaller} \\
\midrule
Source Count & 54 & 17 & 9\\
Attackable Source Count & 5 & 1 & 1\\
Source-to-Sink Flow Count (before) & 116 & 19 & 14\\
Source-to-Sink Flow Count (after) & 5 &  1 & 1 \\
Hijacking Positive & 3 & 0 & 0\\
True Positive Hijacking & 1 & 0 & 0\\
Path Traversal Positive & 2 & 1 & 0\\
True Positive Path Traversal & 2 & 1 & 0\\
Luring Positive & 0 & 0 & 1\\
True Positive Luring & 0 & 0 & 1\\
\bottomrule
\end{tabular}
\end{threeparttable}
}
\end{table}

\paragraph{Path Traversal Vulnerability Case Study} We discovered a zero-day path traversal vulnerability that allows a third-party app (attacker) to gain full read/write access to files private to the wallpaper platform app, which is a LV2 app. The vulnerable component is the content provider {\em WallpaperProvider}, which provides an interface for other applications to request files from it. Our analysis tool showed that this component is exported and not protected by any Android permission. We verified that a control flow exists from the external source (uri of line 1 of Listing~\ref{lst:wallpaper}) to the file operation sink at line 9 {\em ParcelFileDescriptor.open()} and that tainted data reached this sensitive file operation. Our symbolic execution engine collected the relevant path constraints, which require the incoming URI to begin with {\em com.oneplus.wallpaper/image}. Given that this is a path traversal vulnerability, the file constraint suggests providing an app-private file for the initial input generation. With this information, we were able to produce the correct input and maliciously modify wallpaper app's private configuration files from the context of untrusted app.

\vspace{0.1in}
\begin{lstlisting}[language=Java, caption=Vulnerable function openFile in the WallPaper app, label=lst:wallpaper]
public ParcelFileDescriptor openFile(Uri uri, String str) throws FileNotFoundException {
        if (sUriMatcher.match(uri) != 10) {
            return;
        }
        //Omitted code
        ...
        File imageDir = WallpaperContract.Images.getImageDir(context);
        if (imageDir.exists() || imageDir.mkdirs()) {
            return ParcelFileDescriptor.open(new File(imageDir, uri.getLastPathSegment()), ParcelFileDescriptor.parseMode(str));
        }
        ...
\end{lstlisting}
\vspace{0.1in}

The root cause of this vulnerability lies in the careless usage of {\em uri.getLastPathSegments()}, which obtains the last segment of an adversary-controlled URI. Developers often assume that the last segment of a URI will simply be the filename. However, an adversary can exploit this assumption by using {\em ../} to perform a path traversal attack, effectively navigating to parent directories and accessing files outside the intended scope. To defend against this attack, we recommend adding sink sanitization using the getCanonicalPath method, which resolves such path traversal attempts and returns the canonical pathname, together with condition checks. 

\vspace{10pt}
\begin{lstlisting}[language=Java, caption=Vulnerable function openFile in the ClientProvider component of the SecTelephony app, label=lst:clientProvider]
public ParcelFileDescriptor openFile(Uri uri, String str) {
    //Omitted code
    ...
    ...
    Uri.Builder buildUpon = uri.buildUpon();
    String path = uri.getPath();
    if (!(buildUpon == null || buildUpon.build().getQueryParameter("encode") == null || !buildUpon.build().getQueryParameter("encode").equals("path"))) {
        path = buildUpon.build().getEncodedPath();
    }
    
    ...
    //Omitted Code
    LOG.d(str2, "openFile: uri: " + path);
    File file = new File(path);
    ... 
    //Omitted Code
    return ParcelFileDescriptor.open(new File(path), i);
}
\end{lstlisting}
\vspace{10pt}

\paragraph{Path Traversal Vulnerability on Guarded Component} In addition to the path traversal vulnerability discussed above, we discovered two instances of path traversal  vulnerabilities that require a more privileged attacker.  These vulnerabilities are characterized by the fact that they require the adversary to declare certain Android permissions in order to successfully carry out the attack (i.e., a entry constraint must be satisfied).

\sloppypar \tool identified a zero-day vulnerability in Samsung's SecTelephony app's ClientProvider component, which allows adversaries to read/write files accessible to the app (Listing~\ref{lst:clientProvider}). Data flow analysis confirmed that the adversary-controlled uri reaches the ParcelFileDescriptor.open() sink (line 17) through two distinct paths: getPath() (line 6) and getEncodedPath() (line 8). Symbolic execution showed both paths are satisfiable, increasing the vulnerability's impact and complicating mitigation.

Access control policy analysis revealed that exploiting this vulnerability requires a signature-level Android permission, typically reserved for LV3 apps. However, we discovered multiple LV2 OEM apps, including Samsung Browser, possessing the necessary permissions. This highlights the complex permission interactions in the Android ecosystem and expands the potential attack surface.

All other path traversal positives found on Samsung are guarded by custom permissions. While not considered exploitable since they require signature level permission, the reported positives still represent security risks. Notably, Samsung has begun addressing these issues in Android 14, adding checks to at least one vulnerable ContentProvider, validating \tool's findings.

\paragraph{Luring Vulnerability Case Study} 
\tool discovered a luring vulnerability in the GameInstaller platform app (a OnePlus OEM application). This vulnerability allows an attacker to provide an installation APK file to the victim. The vulnerable component, \texttt{InstallProvider}, is believed to be designed to provide installation services for the game center app. \tool successfully generated input that drives the program to the Java file open operation and verified that GameInstaller operates on the adversary-chosen file to which the app is directed in the \texttt{SilentInstall} function. 

However, we did not observe the provided malicious APK being installed on the device.
Due to heavy obfuscation, manual inspection of the code proved challenging, as conventional tools like JADX failed to decompile the code block responsible for APK installation. We suspect that the malicious APK needs to be crafted in a specific way to successfully exploit this vulnerability. Unfortunately, this is one of the current limitations of our tool; we do not collect constraints to help craft malicious file payloads or specify what malicious content must be present in the file.

Although we were unable to fully validate the vulnerability is exploitable, it is important to note that the luring payload was automatically generated by \tool and successful in getting the victim to open the targeted file. Currently, the only protection in place requires a third-party application to declare a custom Android permission, \texttt{oneplus.permission.SILENT\_INSTALL}, which is a normal-level Android permission that is granted at install time without the user's explicit agreement. This level of protection is insufficient, considering the potential impact of a malicious APK being silently installed on a user's device.

\paragraph{Hijacking Vulnerability}
Our analysis of the cidmanager app (telephony-related) on Samsung Android 14 revealed a critical hijacking vulnerability. Sensitive file operations, including read, write, and chmod, are performed on resources within the /data/log/omc folder, which could potentially be under adversary control. We identified multiple source-to-sink flows with satisfiable path constraints and no sanitization before file operations. However, exploiting this vulnerability requires log permission (e.g., process like adb), which somewhat limits its scope.

In addition to the SecTelephony vulnerability, our tool uncovered other hijacking vulnerabilities related to legacy storage locations in external storage. A notable example is the sdm system app on Samsung Android 14 devices, which reads data files from potentially compromised legacy storage locations, exposing itself to attacks. This finding aligns with previous research~\cite{lee2023polyscope}, which highlighted the significant risks of placing sensitive files in the root directory of external storage. Such practices can enable third-party applications to hijack resources, leading to unauthorized access and data manipulation.

\subsection{Exploit Generation with LLM}
\label{subsec: genLLM}
We conducted a preliminary evaluation of using Large Language Models (LLMs) to generate exploit code for vulnerability validation, focusing on 16 true positive path traversal cases. For 7 simple cases involving single-component vulnerabilities, LLMs successfully generated appropriate test cases when provided with vulnerability descriptions, decompiled code, and vulnerability locations. However, for 9 complex cases involving inter-component communication and partially decompiled code, LLMs initially produced only 1 partially correct test case. By incorporating our static analysis results—including correct program entry points, component sequencing, and payload structures—we enabled LLMs to generate usable test cases for all 9 complex scenarios. While promising, these results are based on a limited sample and warrant further investigation to fully assess the approach's potential and limitations in mobile app security testing.

\subsection{OEM Improvement in Defense}
\label{subsec:improv}
Our analysis reveals significant enhancements in Samsung's filesystem security between Android 12 and 14. The use of getCanonicalPath for path sanitization increased from 4 instances in Android 12 to 27 in Android 14, demonstrating improved defense against path traversal attacks. Additionally, we observed a substantial reduction in the use of vulnerable legacy storage locations. These improvements likely account for the decreased number of filesystem vulnerabilities in Samsung 14 devices compared to Android 12, as shown in Table \ref{tab:vuln_overview}. This reduction shows Samsung's growing awareness of path traversal vulnerabilities and their commitment to enhancing the security of their Android implementation.

\section{Discussion}
\label{sec:discussion}
In this section, we review limitations in the \tool approach and examine potential improvements for future work.
\subsection{Analysis Accuracy}
\label{subsec:accuracy}
\paragraph{False Positives} 
During our analysis, we encountered instances where \tool reported potential vulnerabilities that were later identified as false positives. These false positives arise due to the path constraints collected by the tool being impossible to satisfy at runtime. One such example was observed in a hijacking vulnerability case, where the collected constraint required the static variable \texttt{loglevel} to be greater than 3. However, during our verification process, we discovered that the static variable was initialized to 2, and there was no way to modify its value anywhere in the program.
The reason behind this incorrect constraint lies in the limitations of our PDG construction process. In this particular case, we did not include class initialization during the PDG construction, leading to an incomplete representation of the program's behavior. As a result, \tool generated a path constraint that was infeasible in practice, as it did not account for the initial value of the static variable.

Other potential factors contributing to false positives are the lack of precise information about the runtime environment and invocation of native code with through JNI. Static analysis tools rely on an abstract representation of the program and may not have access to all the necessary contextual information. This limitation can result in the generation of path constraints that are feasible within the abstract model but not in the actual runtime environment. We might be able to use concolic technique similar to~\cite{tiro, centaur, condroid, CATE}, obtaining environmental context through concrete execution. Furthermore, to improve the handling of native code, one promising direction would be to integrate tools like Angr~\cite{angr}, which specializes in handling native code.

\paragraph{False Negatives}
While \tool aims to provide comprehensive vulnerability detection, it's important to acknowledge potential sources of false negatives. Under normal circumstances, our analysis over-approximates program behavior, ensuring soundness. However, we rely on Soot and FlowDroid for lifting Android DEX files into Soot's intermediate representation (IR) and constructing the Program Dependence Graph (PDG). In some cases, these tools encounter errors during this process, resulting in an incomplete PDG. This incompleteness can lead to potential false negatives, as certain control and data flows may be missing from our analysis. Verifying these false negatives is challenging, as it requires manual inspection of Smali code to reconstruct the actual control and data flows. 

Another challenge is the presence of reflection and dynamic component registration in Android applications. Reflection allows developers to invoke methods and access fields dynamically at runtime, making it difficult for static analysis tools to determine the exact behavior of the program. Similarly, dynamic component registration enables the creation and registration of components at runtime, which can be challenging to capture accurately in the PDG. These dynamic features can result in missing or incomplete information in our static analysis, potentially leading to undetected vulnerabilities. To address these challenges more effectively, future work should focus on improving the Flowdroid tool and better handle reflection and dynamic component registration.

\paragraph{Assessing Exploitability}
While \tool is effective at generating payloads to drive the victim to open adversary-chosen files via path traversal, hijacking, and luring, it is crucial to note that programs may not use these adversary-chosen resources in a manner that causes exploitation. 
If a program validates the content of the adversary-chosen file, they may still avoid exploitation.  
For example, the luring vulnerability found in Section~\ref{subsec:case_study} causes the victim to open an adversary-chosen APK, but we have not yet seen it loaded.  While we may still find a way to install our APK, it may be that some validation is performed that the generated payload fails to pass.  
It would valuable to conduct further analysis to assess the exploitability of identified vulnerabilities. This analysis should take into account how the adversary-chosen resource's data is used to possibly cause the leakage of private data and/or the illicit modification of program and/or file data.  

\paragraph{Improve On-device Testing}
While \tool, combined with LLM-based code generation, can produce test code to evaluate potential filesystem vulnerabilities, the process of installing the test app and conducting on-device testing remains semi-automatic. To further streamline the testing process and scale up vulnerability validation across a wider range of OEMs, full automation of these final steps is necessary. Future work could explore integrating automated app installation and test execution frameworks, potentially leveraging techniques from automated testing tools like Intent Fuzzer~\cite{intent-fuzzer} to achieve end-to-end automation of the vulnerability validation process or instrument Android's IPC framework, similar to approach done in IntelliDroid~\cite{intellidroid}, to inject inputs automatically.

\section{Related Work}
\label{sec:related_work}
Researchers have long studied filesystem vulnerabilities, particularly those involving name resolution and TOCTTOU attacks~\cite{mcphee74,bishop-dilger}. Various defensive strategies have been proposed, ranging from application-level modifications~\cite{raceguard,27,dean-hu,tsafir} to kernel-level enhancements~\cite{chapin_tocttou,28,venema_ndss_2010,openwall,ty-race,35}. These approaches focus on maintaining file access invariants~\cite{raceguard,chapin_tocttou,35,27,28,ty-race}, preserving namespace invariants~\cite{venema_ndss_2010,openwall}, or providing safe access methods~\cite{dean-hu,tsafir}. However, balancing program-level visibility with system-level knowledge remains challenging~\cite{johnson-tocttou}.

Attack surface identification in Android systems has been extensively studied through access control policy analysis~\cite{lee21usenix, lee2023polyscope, chen17acsac, jaeger02sacmat,setools, enck09ccs,wae+17,Wang2015, BigMAC}. While these approaches effectively highlight potential risks in Android customizations, they often fall short of identifying exploitable vulnerabilities.

Directed symbolic execution has proven valuable in analyzing Android applications~\cite{TASMAN, appIntent, parvez16}, addressing issues such as detecting logic bombs~\cite{triggerscope} and code obfuscation~\cite{tiro}. Our work extends these techniques, focusing specifically on filesystem vulnerabilities and introducing novel enhancements for handling Android-specific API calls.

\section{Conclusion}
In conclusion, \tool represents a significant advancement in the static detection of filesystem vulnerabilities in Android systems. By effectively combining access control policy analysis and program analysis techniques, \tool identifies potential attack surfaces and generates inputs to verify vulnerabilities. The incorporation of file constraints derived from access control analysis significantly improves testing efficiency by reducing the number of paths that need to be analyzed. Evaluation results demonstrate \tool's effectiveness, uncovering critical vulnerabilities such as a zero-day in the Wallpaper app that allows adversaries to gain unauthorized access to private files.

\bibliographystyle{IEEEtran}
\bibliography{citations}

\end{document}